\documentclass{article}
\usepackage{amsmath}
\usepackage[margin = 1in]{geometry}
\usepackage[utf8]{inputenc}
\usepackage[compat = 1.1.0]{tikz-feynman}
\usepackage{adjustbox}
\usepackage{graphicx}
\usepackage{caption}
\usepackage{float}
\usepackage{slashed}
\usepackage{authblk}
\usepackage{multirow}
\usepackage{color}
\usepackage{colortbl}
\usepackage{array}
\usepackage{booktabs}
\usepackage[colorlinks=true]{hyperref}
\usepackage{cite}

\title{New energy spectra in neutrino and photon detectors to reveal hidden dark matter signals}
\author[1,2]{Wim Beenakker \thanks{w.beenakker@science.ru.nl}}
\author[2,3]{Sascha Caron \thanks{scaron@nikhef.nl}}
\author[1]{Jochem Kip \thanks{jochem.kip@ru.nl}}
\author[4]{Roberto Ruiz de Austri \thanks{rruiz@ific.uv.es}}
\author[1,3]{Zhongyi Zhang \thanks{zzhang@nikhef.nl}}
\affil[1]{Institute for Mathematics, Astrophysics and Particle Physics, Radboud University Nijmegen, Heyendaalseweg 135, Nijmegen, the Netherlands}
\affil[2]{Institute of Physics, University of Amsterdam, Science Park 904, 1018 XE Amsterdam, The Netherlands}
\affil[3]{Nikhef, Science Park, Amsterdam, The Netherlands}
\affil[4]{Instituto de Física Corpuscular, CSIC-Universitat de València, E-46980 Paterna, Valencia, Spain}
\newcommand{\bpm}{\begin{pmatrix}}
\newcommand{\epm}{\end{pmatrix}}
\newcolumntype{?}{!{\vrule width 1pt}}
\definecolor{Gray}{gray}{0.85}

\begin{document}

\maketitle
\begin{abstract}
Neutral particles capable of travelling cosmic distances from a source to detectors on Earth are limited to photons and neutrinos. Examination of the Dark Matter annihilation/decay spectra for these particles reveals the presence of continuum spectra (e.g. due to fragmentation and W or Z decay) and peaks (due to direct annihilations/decays). However, when one explores extensions of the Standard Model (BSM), unexplored spectra emerge that differ significantly from those of the Standard Model (SM) for both neutrinos and photons. In this paper, we argue for the inclusion of important spectra that include peaks as well as previously largely unexplored entities such as boxes and combinations of box, peak and continuum decay spectra.
\end{abstract}

%%%%%%%%%%%%%%%%%%%%%%%
\section{Introduction}
%%%%%%%%%%%%%%%%%%%%%%%%
The search for Dark Matter (DM) by indirect detection is the subject of many studies. A large number of experiments have investigated the cosmic antiproton, positron, photon and neutrino spectra. Notable experiments include but are not limited to, AMS-02~\cite{AMS-02:AGUILAR20211}, Fermi-LAT~\cite{FERMI-LAT:Ackermann_2015}, Icecube~\cite{ICECUBE:https://doi.org/10.48550/arxiv.2107.11224}, ANTARES~\cite{ANTARES:Albert_2020}, H.E.S.S.~\cite{HESS:Abdalla_2022}, Pierre Auger~\cite{Pierre_Auger:Aloisio_2023}, and VERITAS~\cite{VERITAS:zitzer2017veritas} which have been measuring charged cosmic rays, gamma rays, and neutrinos for decades. With the upcoming construction of new experiments such as KM3Net~\cite{KM3Net_proj_limit:Adri_n_Mart_nez_2016}, the CTA observatory~\cite{CTA:2018}, and GRAND~\cite{GRAND:Martineau_Huynh_2019} the sensitivity to potential neutral particles arising from DM annihilation or decay will increase significantly.\\
Historically the investigated particle spectra resulting from annihilating or decaying DM have focussed on processes involving two Standard Model (SM) final state particles, e.g. DM DM $\to b\bar{b}$, DM DM $\to \tau^+ \tau^-$, etc, which subsequently undergo decay, fragmentation and hadronization to produce antiproton, positron, photon, and neutrino spectra. These spectra are produced by well-understood mechanisms, leading to detailed analyses that can produce strong limits on the properties of DM. Some notable examples are the potential AMS-02 antiproton excess and the Fermi LAT gamma-ray excess, both of which have been explained by DM annihilation with a DM mass in the $\mathcal{O}$(100) GeV region~\cite{AMS_02_DM:Cui_2018, AMS-02_DM:Cholis_2019, AMS-02_DM:Cui_2017, AMS-02_DM:Cuoco_2017, AMS-02_DM:Cuoco_2017_2, AMS-02_DM:Cuoco_2019, AMS-02_DM:Lin_2019, AMS-02_DM:Reinert_2018, FERMI_DM:Achterberg_2015, FERMI_DM:Calore_2015, FERMI_DM:Daylan_2016, FERMI_DM:Gordon_2013, FERMI_DM:Hooper_2011, FERMI_DM:Hooper_2011_2, FERMI_DM:https://doi.org/10.48550/arxiv.0910.2998, FERMI_DM:https://doi.org/10.48550/arxiv.0912.3828, FERMI_DM:https://doi.org/10.48550/arxiv.1709.10429}. However, if one considers possible extensions beyond the standard model (BSM), other, still largely unexplored spectra are possible, which may differ considerably from the known standard spectra.\\
In this paper, we describe a new type of neutrino and photon spectrum in a largely model-independent way. It contains a combination of a well-defined peak, a box and a combination of neutrino/photon spectra produced by the decay, fragmentation and hadronization of SM particles. Of course, a clearly defined peak has been investigated by Icecube~\cite{Icecube_neutrino_line:theicecubecollaboration2023search}, ANTARES~\cite{ANTARES_neutrino_line:Albert_2017} and Fermi-LAT~\cite{Fermi_LAT_gamma_line:Ackermann_2015} among others, as it is easily obtained by DM particles directly annihilating or decaying into the relevant final state resulting in a comparatively clean signal. However, a box shape and, of course, the combination of all three features leads to a significantly different spectrum. These types of spectra have been largely overlooked in both experimental and phenomenological research. To facilitate the search for these spectra, we provide a code to obtain a user-defined non-standard neutrino or photon spectrum by specifying the appropriate parameters.\\
This paper is structured as follows. First, we detail the physics of the non-standard spectra and provide the relevant expressions of the kinematics. Next, some example BSM models are given that can produce non-standard neutrino or photon spectra. Then we detail the working of the sampling code and its verification, after which we provide some elementary parameter sets that capture the most important features of the spectra in order to facilitate experimental searches. We end with our conclusions.

%%%%%%%%%%%%%%%%%%%%%%%%%%%%%%%%
\section{Theoretical background}
%%%%%%%%%%%%%%%%%%%%%%%%%%%%%%%%%%
There are only two types of neutral particles that can travel cosmic distances from a source to detectors on or near Earth, namely photons and neutrinos. In the following subsections, the kinematics of the spectrum of a particle will be discussed as model-independently as possible. Both neutral particles can be described by the same kinematics, since they are both massless or have negligible mass. The main differences, of course, lie in the possible models that can produce such spectra. However, no such assumptions will be made in the following subsections.\\
Moreover, both DM annihilation and DM decay can produce cosmic rays. The kinematics for DM decay is identical to DM annihilation, with the only difference being that the initial energy for (non-relativistic) DM decay is $M_{\rm DM }$, while for DM annihilation it is $2M_{\rm DM }$. Thus, to obtain the kinematic expressions for the DM decay from those for the DM annihilation, one simply has to replace $M_{\rm DM }$ by $\frac{1}{2}M_{\rm DM }$\\
In the following subsections, we assume the standard scenario that two DM particles annihilate to neutrinos in order to simplify the discussion.

%%%%%%%%%%%%%%%%%%%%%
\subsection{The box}
%%%%%%%%%%%%%%%%%%%%
The simplest non-standard spectrum arises from two DM particles annihilating into two BSM particles, $X$, which subsequently decay into neutrinos:
\begin{align}
    {\rm DM} \, {\rm DM} \to XX && X \to \nu \nu\,,
\end{align}
where $\nu$ is any SM neutrino. More complicated decays of $X$ are of course possible, which will be discussed in section \ref{subsec:Xdecays}.  Depending on the model, of course, the $XX$-pair can also be a $X\overline{X}$-pair, and any $\nu$ may well be an $\overline{\nu}$. This kind of DM annihilation leads to a 'box' shape of the neutrino spectrum. In the rest frame of the particle, $X$ both neutrinos produced by the decay of $X$ have a clearly defined energy of $M_X/2$, where $M_X$ is the mass of $X$. However, the neutrinos need to be Lorentz boosted into the center-of-mass frame of the annihilating DM particles as they are now evaluated in the rest frame of the particle $X$, which of course differs from the center-of-mass frame. Assuming that ${\rm DM}$ has zero momentum at annihilation, using spherical coordinates in the rest frame of $X$, and choosing the motion of $X$ in the $z$ direction and thus boosted in the $z$ direction, the energy of the neutrinos results in:
\begin{align}
        E_{\text{box}} = \frac{M_X}{2}\left(\cosh{(\eta)} + \sinh{(\eta)}\cos{(\theta)}\right)  && \eta = \cosh^{-1}{\left(\frac{M_{\rm DM}}{M_X}\right)}\,.
\end{align}
Here $M_{\rm DM}$ is the DM mass. The $\cos(\theta)$ term is due to the spherical coordinates in the rest frame of $X$. To obtain a uniform distribution of points on a sphere, i.e. an isotropic decay, $\cos(\theta)$ must be sampled uniformly between 0 and 1, as opposed to sampling $\theta$ uniformly between 0 and $2\pi$\footnote{See \url{http://corysimon.github.io/articles/uniformdistn-on-sphere/} for an in-depth explanation.}. This uniform distribution in $\cos(\theta)$ results in the neutrino having an equal probability of having any energy within the bounds of $\cos(\theta) = -1$ and $+1$, thereby resulting in a flat 'box' shape. \\
However, a particle $X$ will in general not decay isotropically if, for example, it is polarized, or has an asymmetric coupling to different helicities~\cite{Box_differ:Garcia_Cely_2016}. The probabilty amplitude for the decay of $X$ must be proportional to $\langle m^\prime, S| R(\theta) |m, S\rangle$, with $|m, S \rangle$ and $|m^\prime, S \rangle$ the initial and final state respectively, where $S$ is the total spin of $X$, $m$ the angular momentum of $X$ in its flight direction, $m^\prime$ the helicity diffence between the decay products of $X$, and $R(\theta)$ a rotation operator. The complete results for various spin and helicity configurations are provided in~\cite{Box_differ:Garcia_Cely_2016}, which we have implemented here. The shapes of the boxes are given by:
\begin{align}
    &\left.\frac{dN_{\text{box}}}{dE}\right|_{S=0, m = 0} &&\propto 1\,, \nonumber\\
    &\left.\frac{dN_{\text{box}}}{dE}\right|_{S=1/2, m = \pm 1/2} &&\propto 1 \pm (C_{1/2} - C_{-1/2})\frac{2E-E_X}{\sqrt{E_X^2 -M_X^2}}\,,\nonumber\\
    &\left.\frac{dN_{\text{box}}}{dE}\right|_{S=1, m = 0} &&\propto 4EE_X - 4E^2 - M_X^2\,,\nonumber\\
    &\left.\frac{dN_{\text{box}}}{dE}\right|_{S=1, m = \pm 1} &&\propto E_X^2 - 2 E E_X + 2E^2 - \frac{1}{2}M_X^2 \pm (C_1-C_{-1})(2E-E_X)\sqrt{E_X^2-M_X^2}\,. \label{eq:box shapes}
\end{align}
Here $E_X$ is the energy of $X$ in the CM frame, $C_{m^\prime}$ are model-dependent positive normalized coefficients, $\sum_{m^\prime} C_{m^\prime} = 1$, that determine how $X$ couples to the various polarizations. Note that for $S=1/2$ the flat box is regained for either $C_{1/2} = C_{-1/2}$, or when $X$ has no preferred helicity. For a vector $X$ the flat box is regained when $X$ is unpolarized.

%%%%%%%%%%%%%%%%%%%%%%
\subsection{The peak}
%%%%%%%%%%%%%%%%%%%%%%
Another process is the annihilation of two DM particles into a neutrino and a BSM particle $X$:
\begin{align}
    {\rm DM} {\rm DM} \to \nu X\,.
\end{align}
The neutrino, which comes directly from DM annihilation, forms a clearly defined peak in the spectrum with an energy of
\begin{align}
    E_{\rm peak}  = \frac{4M_{\rm DM}^2 - M_X^2}{4M_{\rm DM}}\,. \label{eq:simple_peak}
\end{align}
%%%%%%%%%%%%%%%%%%%%%%%%%%%%%%%%%%%%%%%%%%%%%%%%%%%%%%
\subsection{Alternative decay modes for particle $X$}
\label{subsec:Xdecays}
%%%%%%%%%%%%%%%%%%%%%%%%%%%%%%%%%%%%%%%%%%%%%%%%%%%%%%
In most realistic BSM models, a possible particle $X$ does not have a 100\% branching ratio into $\nu \nu$, but can have multiple different decay modes, e.g. ${\rm BR}[X \to \nu Z] = 0.5$ and ${\rm BR}[X \to W^+e^-] =0.5$. The specific decay modes depend, of course, on the details and parameters of the chosen BSM model. Here, the assumption is explicitly made that $X$ can only decay into SM particles. The SM particles\footnote{Naturally except for neutrinos and $e^\pm$.} will undergo fragmentation, hadronization, and decay, thereby also producing neutrinos or radiating off photons. A labelling is made of the neutrinos that come from the various stages of DM annihilation into neutrinos: the neutrinos that form the peak are called primary neutrinos, those in the box are called secondary, and all neutrinos that come from the fragmentation and hadronization of an SM particle are called standard neutrinos:
\begin{align}
    \includegraphics[width=\textwidth]{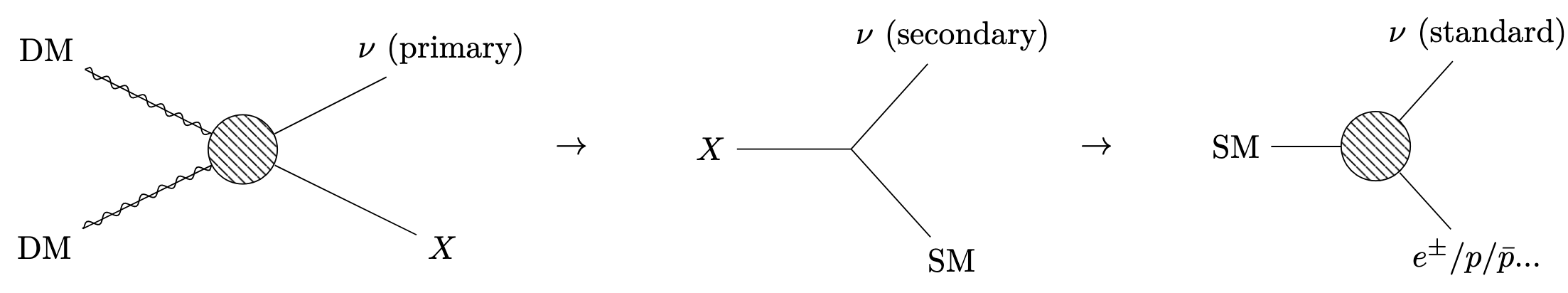} \label{eq: feynman diagrams prim sec prim}
\end{align}
% \begin{align*}
% \adjustbox{valign = c}{
% \feynmandiagram [horizontal = i1 to f1] {
%   i1 [particle = ${\rm DM}$] --  a [blob] -- i2 [particle = ${\rm DM}$],
%   i1 -- [photon] a -- [photon] i2,
%   i1 -- [opacity = 0.0] i2,
%   f1 [particle = $X$] -- a -- f2 [particle = $\nu$ (primary)],
%   f1 -- [opacity = 0.0] f2,
% };
% }
% \rightarrow
% \adjustbox{valign = c}{
% \feynmandiagram[horizontal = a to b]{
%   i1 -- [opacity = 0] a [particle = $X$]-- [opacity = 0] i2,
%   a --  b ,
%   f1 [particle = $\nu$ (secondary)] -- b -- f2 [particle = SM],
% };
% }
% \rightarrow
% \adjustbox{valign = c}{
% \feynmandiagram[horizontal = a to b]{
%   i1 -- [opacity = 0] a [particle = SM]-- [opacity = 0] i2,
%   a --  b [blob],
%   f1 [particle = $\nu$ (standard)] -- b -- f2 [particle = $e^\pm/p/\bar{p}...$],
% };
% }
% \,.
% \end{align*}
The other products of fragmentation, hadronization and decay, such as $e^\pm$, $\bar{p}$, etc., are
ignored here. Of course, the process ${\rm DM}\,{\rm DM} \to XX$ has no primary neutrinos.\\
The energy range of the box depends, of course, on how the particle $X$ is produced and what decay modes it has:
\begin{align}
    &E_{\text{box}} = \frac{M_X^2-M_{\rm SM}^2}{2M_X}\left(\cosh{(\eta)} + \sinh{(\eta)}\cos{(\theta)}\right)\,, \label{eq:plateau_energy}\\
    &\eta = 
    \begin{cases}
    \cosh^{-1}{\left(\frac{4M_{{\rm DM}}^2+M_X^2}{4M_{{\rm DM}}M_X}\right)} & \text{for} \quad  {\rm DM} {\rm DM} \rightarrow X \nu\,, \\
    \cosh^{-1}{\left(\frac{M_{{\rm DM}}}{M_X}\right)} & \text{for} \quad {\rm DM} {\rm DM} \rightarrow X X\,.
    \end{cases}\nonumber
\end{align}
The production mode of $X$ changes the total energy of the particle $X$ at creation and thus $\eta$, while the decay mode changes the energy of the secondary neutrino, giving the term $M_{\text{SM}}^2$. This results in the total width of the box being
\begin{align}
    E_{\text{box},\text{max}} - E_{\text{box},\text{min}} \equiv \Delta E_\text{box} = \frac{M_X^2-M_{\rm SM}^2}{M_X} \sqrt{\Omega^2 -1}\,,\label{eq:box_width} \\
    \Omega = 
    \begin{cases}
        \frac{4M_{{\rm DM}}^2+M_X^2}{4M_{{\rm DM}}M_X}  &\text{for} \quad {\rm DM} {\rm DM} \rightarrow \nu X \, ,\\
        \frac{M_{\rm DM}}{M_X}&\text{for} \quad {\rm DM} {\rm DM} \rightarrow X X \, . \nonumber
    \end{cases}
\end{align}
Notably, from this equation it can be seen that for ${\rm DM \, DM} \to \nu X$ the energy of the peak is equal to the width of the box if $M_{\text{SM}}=0$. In general $E_{\text{peak}} = \Delta E_{\text{box}} / (1 - (M_{\text{SM}}/M_X)^2)$. So for boxes with the same width, the peak is at the same energy, regardless of the energy of the box. Or, put another way, the position of the peak directly provides the width of the box up to the specific decay mode of $X$. \\
Of course, the mass of the particles DM and $X$ can also be expressed by the initial and final energy of the box:
\begin{align}
    &M_X = \sqrt{E_{\text{box},\text{max}}E_{\text{box},\text{min}}}+\sqrt{E_{\text{box},\text{max}}E_{\text{box},\text{min}}+M_{\text{SM}}^2} \, , \\
    &M_{\text{DM}} = 
    \begin{cases}
        \frac{E_{\text{box},\text{max}}M_X^2}{M_X^2 - M_{\text{SM}}^2} & \text{for} \quad {\rm DM} {\rm DM} \rightarrow \nu X \, , \\
        \frac{(E_{\text{box},\text{max}} + E_{\text{box},\text{min}})M_X^2}{M_X^2-M_{\text{SM}}^2} & \text{for} \quad {\rm DM} {\rm DM} \rightarrow X X  \, . \nonumber \\
    \end{cases}
\end{align}
For this expression, the decay mode of $X$ must be fixed to know $M_{\text{SM}}^2$.\\
The standard neutrinos produced by the hadronization and fragmentation of the SM particle can best be determined with programs such as {\tt Pythia}~\cite{Pythia} and are not analytically determinable. However, pre-computed spectra of ${\rm DM} {\rm DM} \to$ SM SM can be used to sample these neutrinos. For ${\rm DM} {\rm DM} \to$ SM SM the energy of the SM particle is simply $ E_{\rm SM} = M_{\rm DM}$. For $X\to$ SM$_1$ SM$_2$, i.e. two different SM particles, the energy of particle SM$_1$ and SM$_2$ is respectively given by:
\begin{align}
    E_{{\rm SM}_1} = \frac{M_X^2 + M_{{\rm SM}_1}^2 - M_{{\rm SM}_2}^2}{2M_X} \,, && E_{{\rm SM}_2} = \frac{M_X^2 -M_{{\rm SM}_1}^2 + M_{{\rm SM}_2}^2}{2M_X}\,.
\end{align}\\
Thus, to obtain the correct neutrino spectrum of SM$_1$, the spectrum of ${\rm DM } {\rm DM } \to {\rm SM }_1 {\rm SM }_1$ is used, where $M_{\rm DM } = E_{{\rm SM }_1}$, and similarly for the neutrino spectrum of SM$_2$. Since this neutrino spectrum of SM$_{1/2}$ is in the rest state of $X$, any neutrino sampled from this spectrum must be Lorentz boosted into the CM frame of annihilating DM particles, identical to the neutrinos coming directly from the decay of $X$. Additionally, when using the pre-computed spectra of ${\rm DM} {\rm DM} \to$ SM$_1$ SM$_1$ to determine the spectrum of a single particle SM$_1$, its spectrum d$N$/d$E$ is overestimated by a factor of 2 and thus needs to be compensated for. Furthermore, any correlations between the annihilation products of the pre-computed spectra are assumed to be negligible~\cite{Half_spectrum1, Half_spectrum2}.\\
The average number of primary and secondary neutrinos per collision can be straightforwardly determined via counting:
\begin{table}[H]
    \centering
    \begin{tabular}{lcr}
       \multicolumn{1}{c}{Process} & \multicolumn{1}{c}{$N_{\rm prim}$} & \multicolumn{1}{c}{$N_{\rm sec}$} \\
        \midrule
        ${\rm DM} {\rm DM} \to \nu X$ &  1 & ${\rm BR}[X \to \nu {\rm SM}$]\\
        ${\rm DM} {\rm DM} \to X X$ &  0 & $2\cdot{\rm BR}[X \to \nu {\rm SM}$]
    \end{tabular}
\caption{The average number of primary (peak) and secondary (box) neutrinos per DM annihilation.}
\label{tab:number of events}
\end{table}
Note that when $X$ decays into $\nu \nu$, the number of secondary neutrinos is doubled and the number of standard neutrinos is zero for that decay mode. The number of standard neutrinos is determined by the integral of their spectrum, $dN/dE$ and therefore varies from case to case.
%%%%%%%%%%%%%%%%%%%%%%%%%
\section{Model examples}
%%%%%%%%%%%%%%%%%%%%%%%%%%
The only requirement to obtain these spectra is a DM particle capable of either self-annihilation or decay into a mediating particle $X$ that couples to neutrinos and/or photons. Of course, there are some limitations on the possible interactions and decay modes of the DM and $X$ particles. For example, if $X$ is a spin $\frac{1}{2}$ particle, then the decay mode $X\to \nu \nu$ is forbidden by conservation of angular momentum, making a pure box shape impossible. Since the shape of the neutrino spectrum is determined by its kinematics, the spin of the DM or the $X$-particle is irrelevant when it comes to the position of the peak or the box; only the possible couplings are affected. \\
Two examples of candidates for a particle $X$ coupling to neutrinos are: $Z^\prime$ bosons, introduced in a $U_{L_\mu -L_\tau}(1)$\cite{LmuLtau1:PhysRevD.43.R22, LmuLtau2:osti_5072375, LmuLtau3:PhysRevD.44.2118, LmuLtau4:PhysRevD.84.075007} gauge extensions of the SM and unstable heavy neutrinos such as those in an inverse seesaw mechanism\cite{Inverse_seesaw1:Ma_2009, Inverse_seesaw2:PhysRevD.34.1642, Inverse_seesaw3:GONZALEZGARCIA1990108}. A heavy neutrino $\nu_H$ and DM particle $\phi_{\rm DM }$ could, for example, have a non-standard muon neutron spectrum via the decay chain $\phi_{\rm DM} \phi_{\rm DM} \to \nu_\mu \nu_H$ with $\nu_H \to \nu_\mu Z$, and the $Z$ boson gives a continuous neutrino spectrum. The number of potential models can also be much larger; additional gauge groups, different neutrino mass mechanisms or Higgs mechanisms specific to neutrinos could all produce peak or box shapes. Of course, the number of different DM candidates is very large. These include particles that are added manually, such as complex scalar DM, or those that arise as a consequence of the theory itself, e.g. neutralinos in supersymmetry. \\
Note that a single DM species annihilating with itself cannot have a neutrino spectrum consisting only of primary and secondary neutrinos, i.e. ${\rm DM DM } \to \nu X$ with $X\to \nu \nu$. The initial state has an even number of fermions, while the final state has an odd number of fermions, which is of course impossible.\footnote{Two different DM particles annihilating would circumvent such a constraint, e.g. a sneutrino and a heavy neutrino as DM candidates could produce such a spectrum.} For photons, of course, there are no such constraints. It is possible that a process such as ${\rm DM DM } \to \nu \nu X$ with $X \to \nu \nu$, but here no peak is formed because the energy of the primary neutrinos and $X$ is not uniquely determined by the two-body phase space. However, a spectrum with a peak and a box is easily obtained via the decay of fermionic DM. For example, a heavy neutrino decaying into a $\nu$ and a scalar, which subsequently decays into two $\nu$'s.\\
In terms of models that might produce non-standard photon spectra, there are a plethora~\cite{Many_Diphoton_1:PhysRevLett.116.150001, Diphoton2:PhysRevLett.116.151802, Diphoton3:PhysRevLett.116.151803, Diphoton4:PhysRevLett.116.151804, Diphoton5:PhysRevLett.116.151805} of models proposed~\cite{Diphoton_models1:Altmannshofer_2016} to explain the 2015 750 GeV di-photon excess in ATLAS and CMS. More specifically~\cite{photon_box:Ibarra_2012} and~\cite{photon_box2:Ibarra_2015} detail some concrete models regarding photon boxes. It should be noted that if $X$ is electrically charged, as is very possible for a particle coupling to photons, the possible energy range of the spectrum is limited by constraints on the mass of $X$, e.g. by LEP searches. This is in contrast to a particle $X$ that couples to neutrinos via the weak force, which can more easily evade experimental searches. Remarkably, $X$ does not have to be electrically charged to couple to photons. The neutral pion $\pi^0$, for example, has a decay mode into two photons. A non-standard photon spectrum could, for example, be made by a pion-like BSM particle $\Pi^0$ and a DM particle $\phi_{\rm DM }$ by $\phi_{\rm DM } \phi_{\rm DM } \to \Pi^0 \Pi^0$ and $\Pi^0 \to \gamma \gamma$.\\
%%%%%%%%%%%%%%%%%%%%%%%%%%%%%
\section{The sampling code}
%%%%%%%%%%%%%%%%%%%%%%%%%%%%%
\subsection{Sampling procedure}
%%%%%%%%%%%%%%%%%%%%%%%%%%%%%%%%
The sampling of the neutrinos is done by sampling a primary, secondary or standard neutrino according to the probability $N_{\rm prim}/N$, $N_{\rm sec}/N$ and $N_{\rm stand}/N$, respectively, where $N = N_{\rm prim} + N_{\rm sec} + N_{\rm stand}$. Here $N_{\rm prim}$, $N_{\rm sec}$, and $N_{\rm stand}$ refer to the number of primary, secondary and standard neutrinos, as defined by the Feynman diagrams in~\eqref{eq: feynman diagrams prim sec prim}. The primary neutrino is always sampled at $E_{\text{peak}}$ since its energy is fixed and therefore has no distribution. The secondary neutrinos are either sampled uniformly, when the box is flat, or by rejection sampling if $X$ has a polarization, with the box shapes given by Eq.~\eqref{eq:box shapes}. The limits of the box are given by Eq.~\eqref{eq:plateau_energy}. The standard neutrinos are sampled from precomputed spectra, and subsequently Lorentz boosted into the CM frame of the annihilating DM particles as described previously. We use the precomputed spectra from~\cite{computed_spectra1:jueid2023impact, computed_spectra2:Amoroso_2019, computed_spectra3:jueid2023strong}. In order to sample from these precomputed spectra, we numerically integrate them in order to obtain their cumulative distribution function. Subsequently, this cumulative distribution is used in order to perform inverse transform sampling. Notably, secondary photons can give rise to additional particles due to QED showering effects~\cite{Pythia}. We take these effects into account by using the aforementioned precomputed spectra when a secondary particle is a photon.\\
%%%%%%%%%%%%%%%%%%%%%%%%%%%%%%%%%%%%%%%
\subsection{Verification of spectra}
%%%%%%%%%%%%%%%%%%%%%%%%%%%%%%%%%%%%%%%%
\begin{figure}[h]
    \centering
    \includegraphics[width = .45\linewidth]{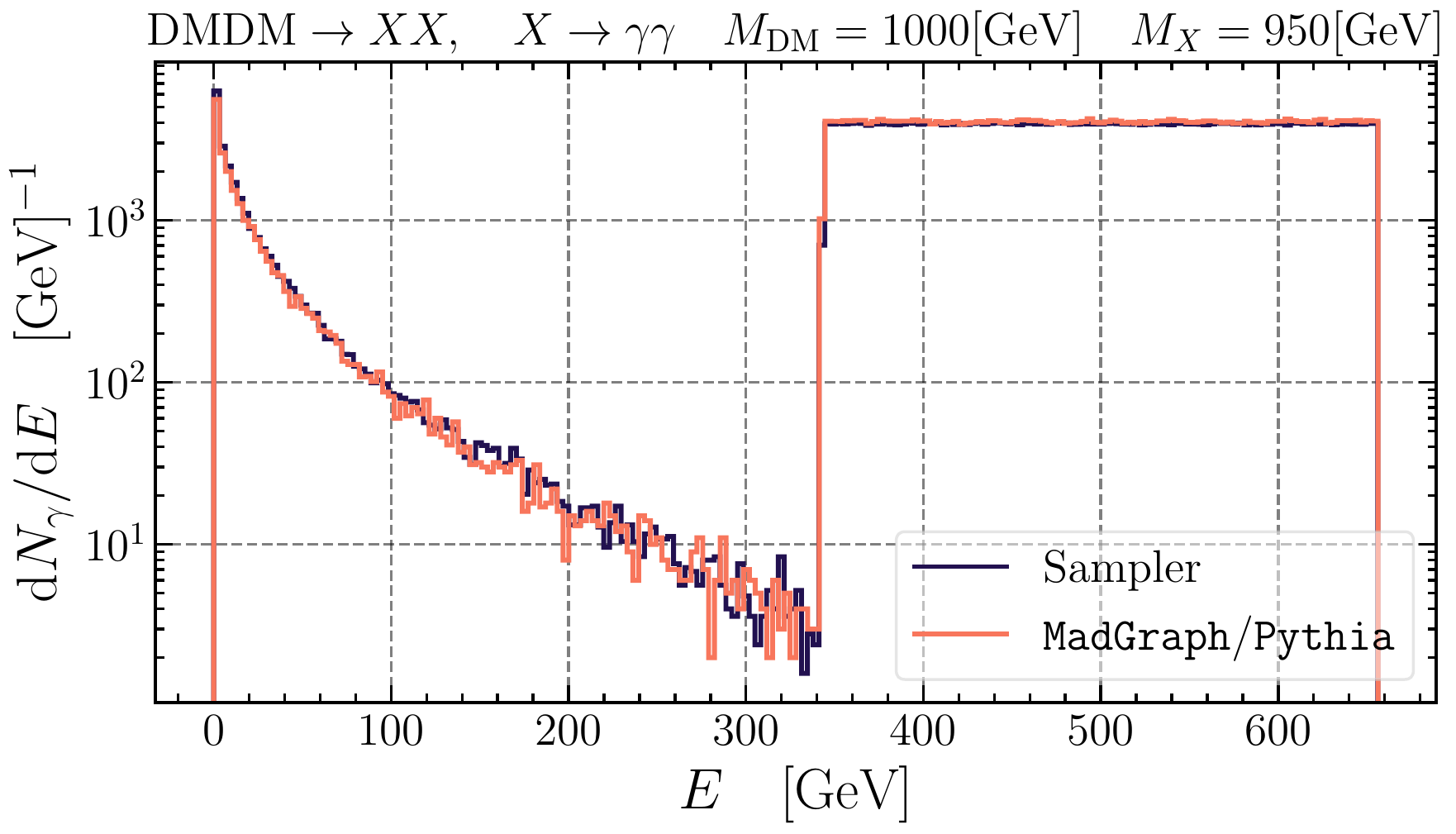}
    \includegraphics[width = .45\linewidth]{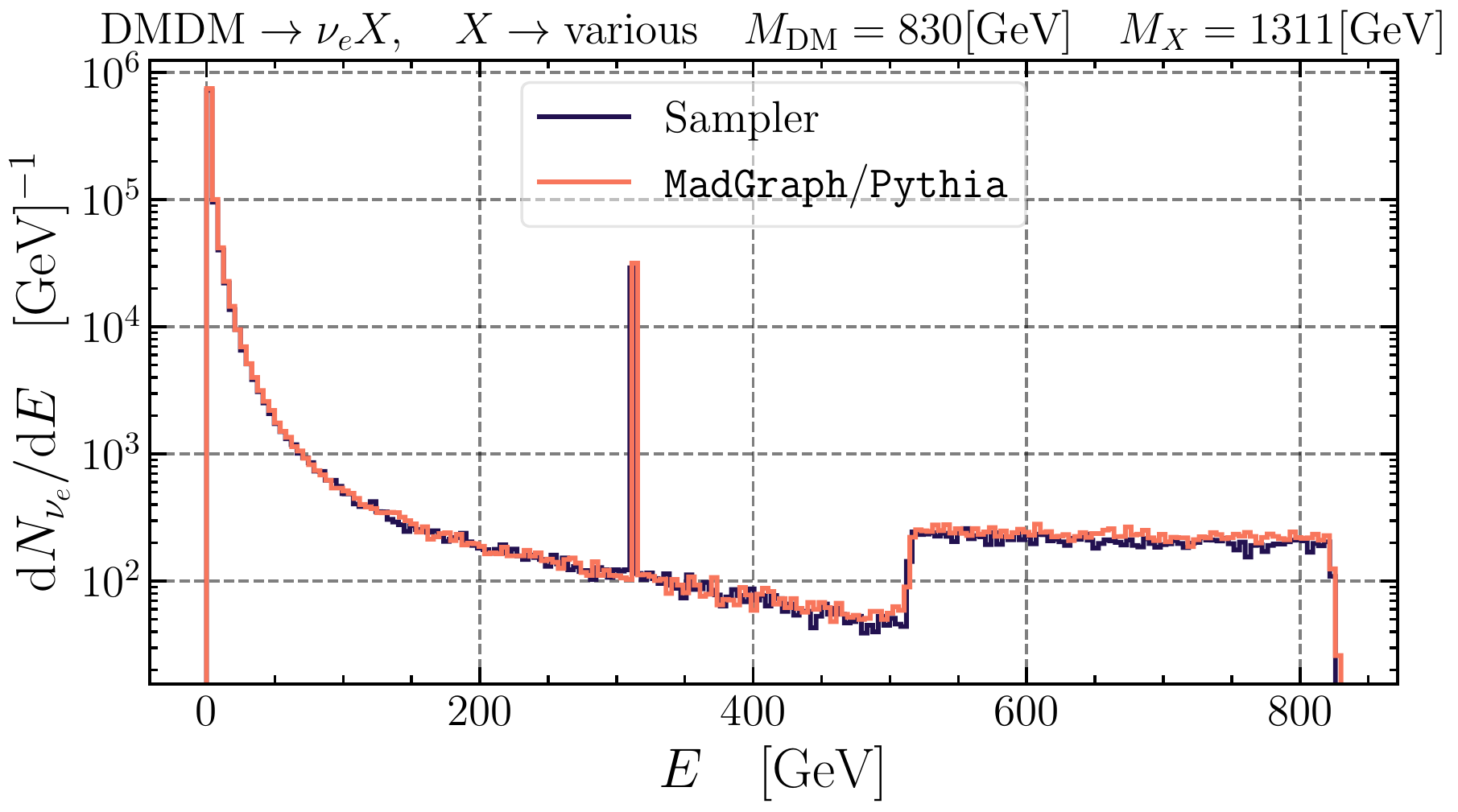}
    \includegraphics[width = .45\linewidth]{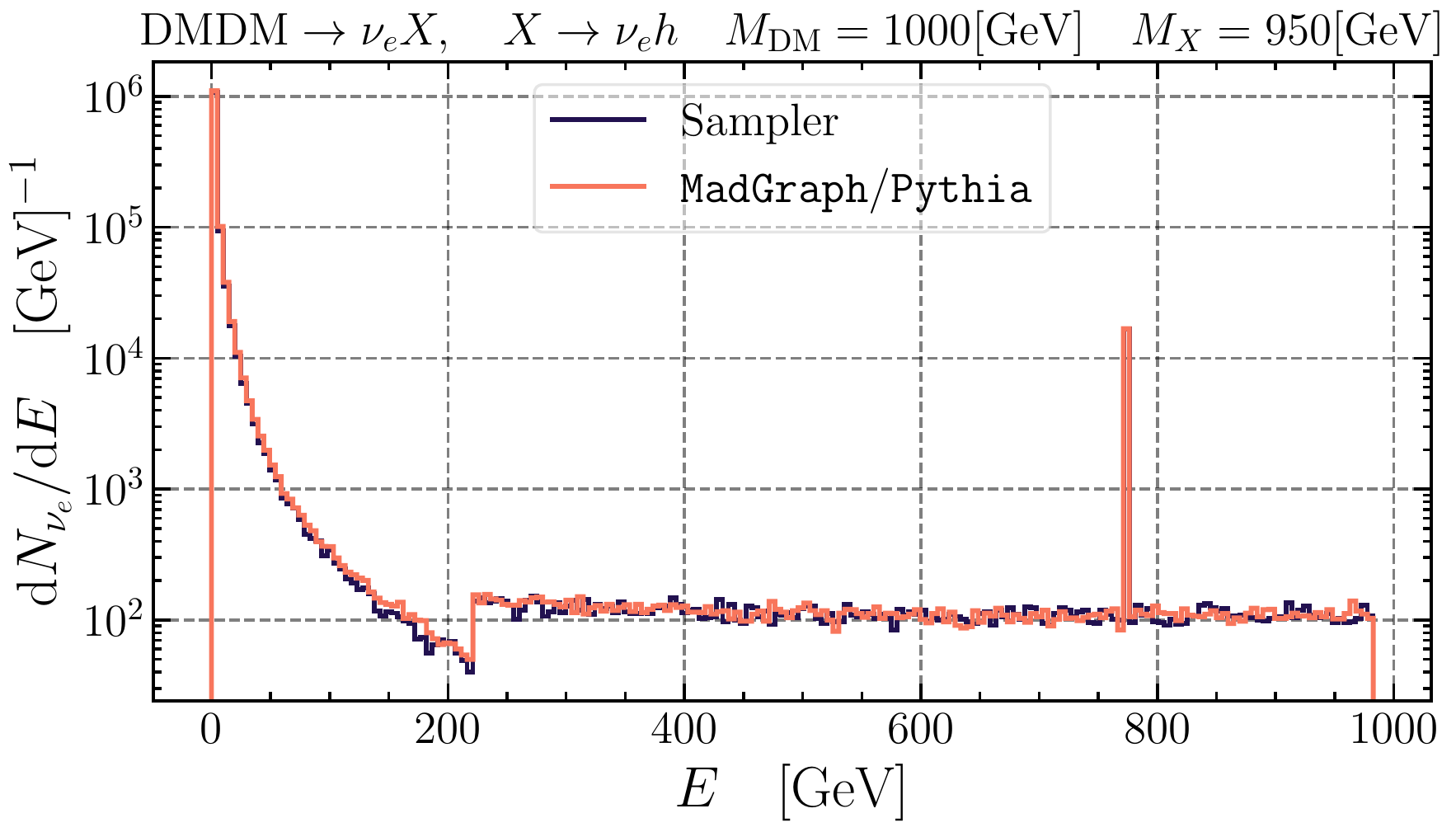}
    \includegraphics[width = .45\linewidth]{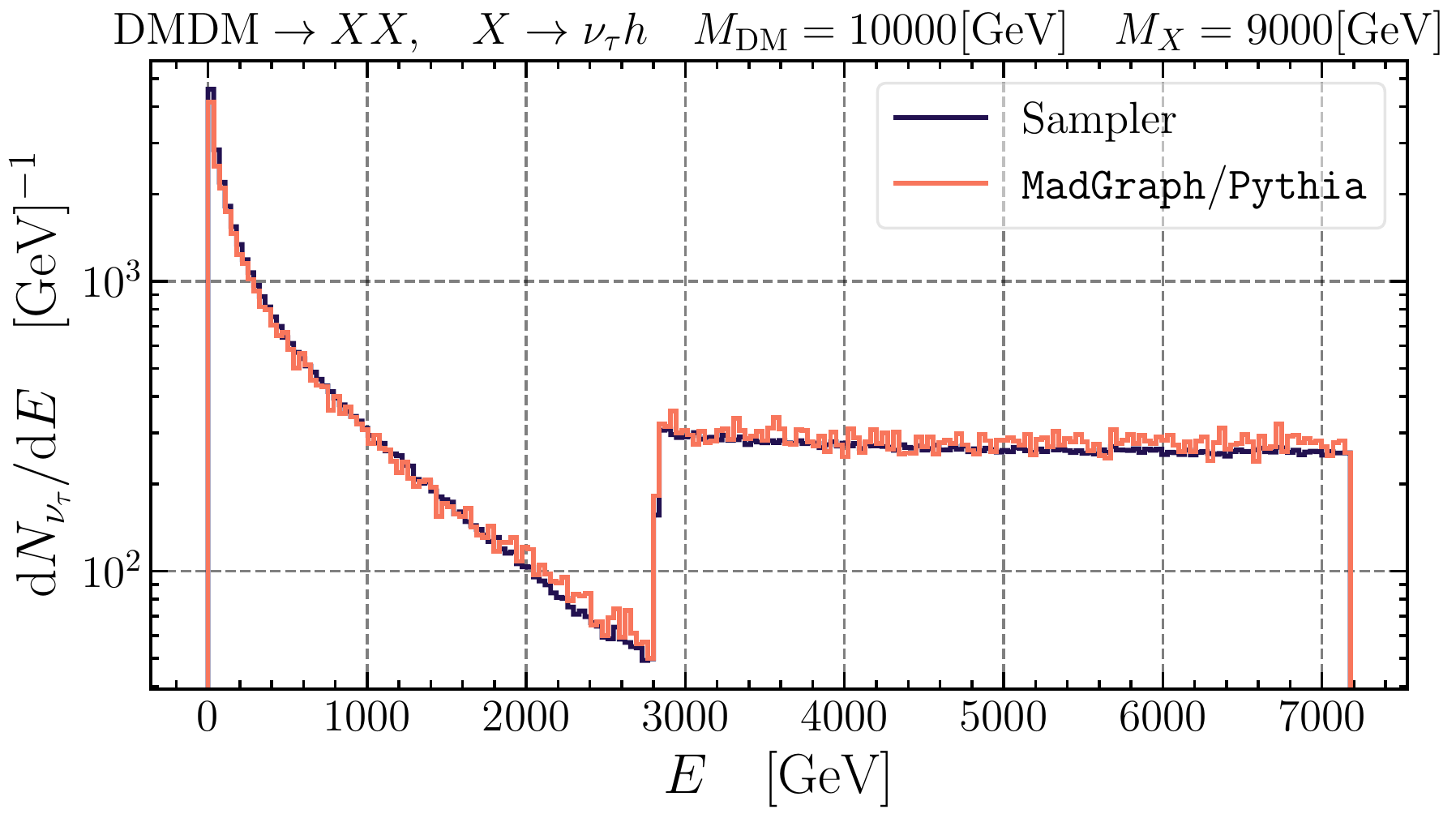}
    \caption{Comparison of the spectra between {\tt MadGraph}/{\tt Pythia} (orange) and the sampling code (blue). The upper left and lower two spectra were generated with simplified models, while the upper right spectrum is an example point from a full model~\cite{BLSSMIS}. The lower two plots show the same model, but evaluated at different energies.}
    \label{fig:comparison}
\end{figure}

\begin{figure}
    \centering
    \includegraphics[width = .45\linewidth]{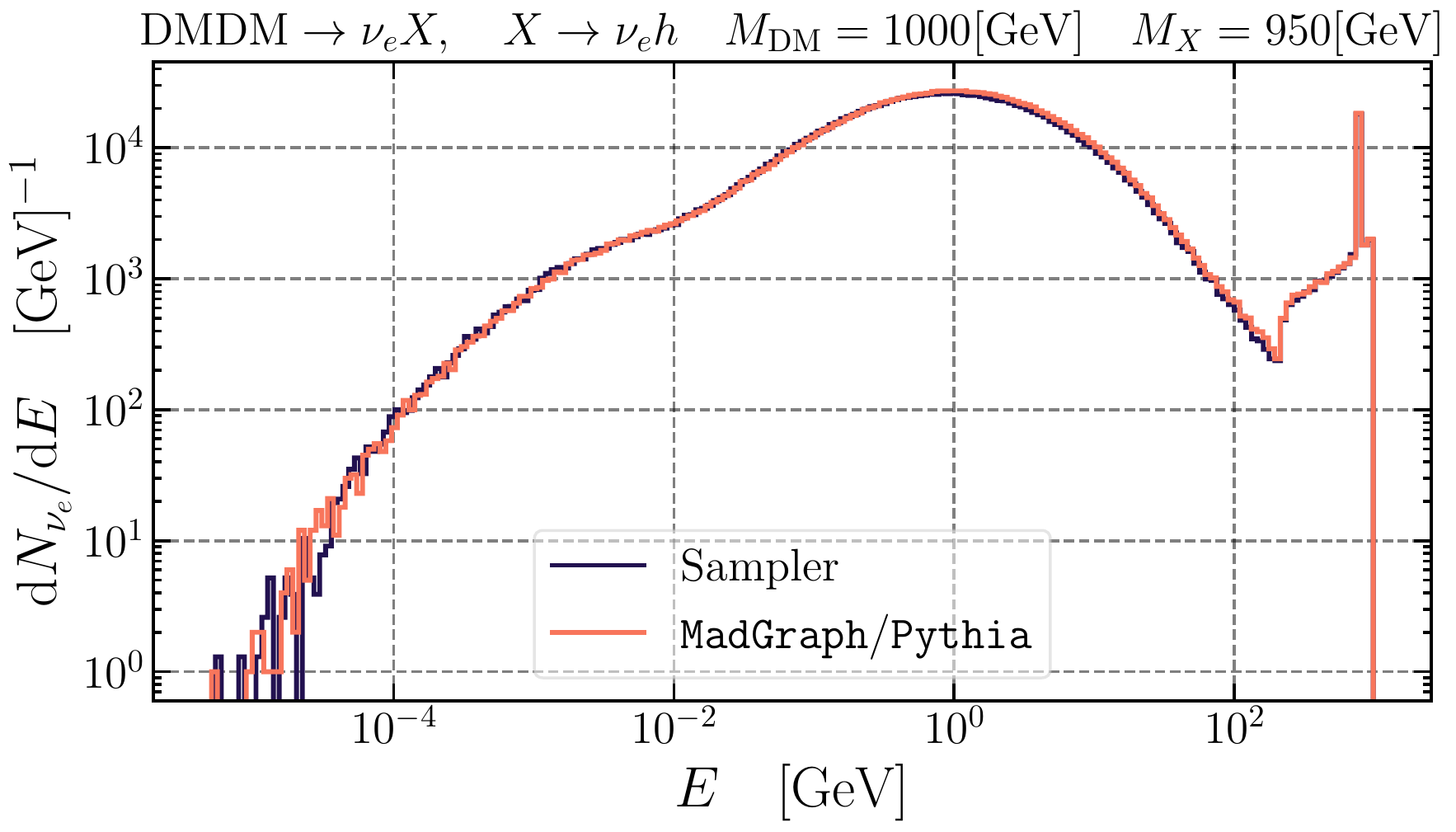}
    \includegraphics[width = .45\linewidth]{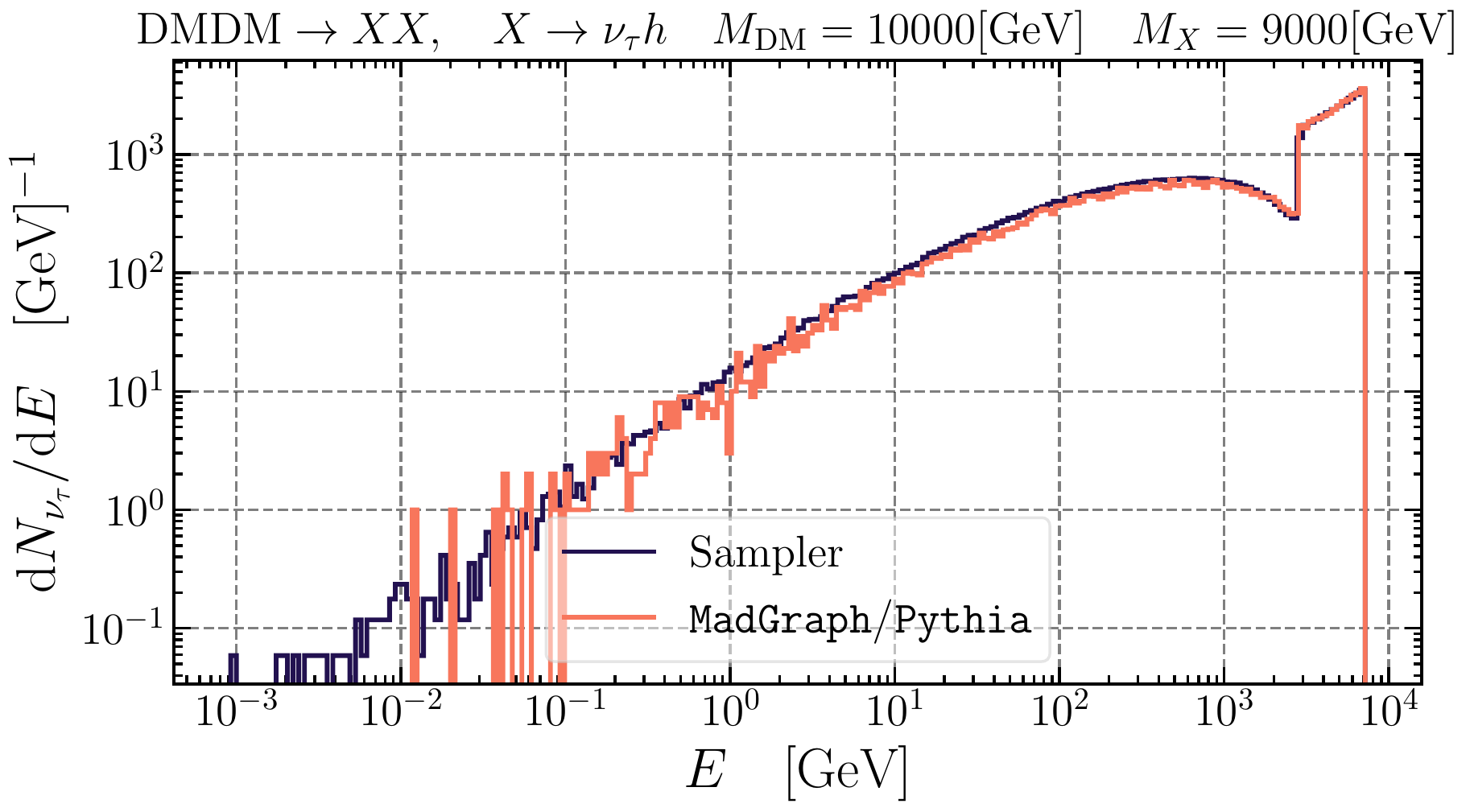}
    \caption{A comparison between the spectra generated by {\tt MadGraph}/{\tt Pythia} (orange) and the sampling code (blue) for two different spectra. The spectra of these plots are the same as the bottom two in \ref{fig:comparison}, but plotted logarithmically. Note that the box now has a slope, which is purely a binning artifact. Since logarithmically equidistant bins are wider for higher values of $E$ in absolute terms, a flat box in linearly equidistant bins will then have a slope when binned logarithmically due to the difference in entries per bin.}
    \label{fig:comparison_log}
\end{figure}
In order to verify the accuracy of this sampler, we perform cross-checks with multiple spectra computed with {\tt MadGraph5} v3.1.1\cite{MadGraph} and {\tt Pythia} 8.309 across a range of masses and decay modes. We deem a simple visual inspection of the spectra to be sufficient to validate the fidelity of the sampler. In figure \ref{fig:comparison} four different example spectra are shown of which the process is indicated in the relevant plot. The top left and bottom two plots are generated through a simplified model in which $X$ is a spin-0 or spin-1/2 particle respectively, while the spectrum from of top right plot is used from~\cite{BLSSMIS}. The decay mode of $X$ of the two spectra from the simplified models both have a branching ratio of 100\% into a single decay mode, while the spectrum in the top right has a more complicated decay mode of ${\rm BR }[X\to \nu h] ={\rm BR }[X\to \nu Z] =1/4$ and ${\rm BR }[X\to W^\pm e^\mp] = {\rm BR }[X\to W^\pm \mu^\mp] = {\rm BR }[X\to W^\pm \tau^\mp] = 1/6$, which shows that the sampler can indeed handle more complicated decay modes. \\
In Figure \ref{fig:comparison_log}, the two lower spectra of Figure \ref{fig:comparison} are plotted on a logarithmic scale, showing that the sampler at low energies performs accurately. There is some erratic behaviour at $E\approx 10^{-4}{\rm GeV}$, but this is simply an artefact of a finite number of Monte Carlo samples.\\
It should be noted that the normalization of the sampler and the spectra produced by {\tt MadGraph} and {\tt Pythia} are different, as the former only samples a spectrum while the latter fully computes an annihilation process. However, this difference can be easily remedied by counting the average number of neutrinos per event, which then directly provides the relative normalization factor. The spectra calculated by {\tt MadGraph} and {\tt Pythia} are the sum of 100,000 iterations, while the sampler's spectra are the result of 1,000,000 samples, which are then normalized.\\
The normalization factor for an arbitrary spectrum can either be determined by counting the number of particles in a peak or box, which is directly related to the average number of particles per DM annihilation/decay as given by Tab.~\ref{tab:number of events}, or can be provided by the sampling code via the {\tt norm\_const()} function. This function provides the normalization factor $(N_{\text{prim}} + N_{\text{sec}} + N_{\text{stand}})/N_{\text{samples}}$. The sampled spectra are not normalized, such that they can easily be used for experimental searches, while the normalization is readily performed for phenomenological purposes.\\
Remarkably, although the spectra shown are $\mathcal{O}$(1-10)TeV, there is no inherent scale for these spectra, since both $M_{\rm DM }$ and $M_X$ are a priori both unconstrained. Similarly, the detectability of any spectrum depends strongly on its annihilation cross-section $\langle \sigma v \rangle$, which is of course model specific and which we do not comment on here. \\
%%%%%%%%%%%%%%%%%%%%%%%%%%%%
\subsection{Required input}
%%%%%%%%%%%%%%%%%%%%%%%%%%%%
The code can sample ${\rm DM} {\rm DM} \to \nu/\gamma X$, ${\rm DM} {\rm DM} \to X X$, ${\rm DM} \to \nu/\gamma X$, or ${\rm DM} \to X X$ in which $X$ can have any decay products and branching ratios into SM particles. The required input is as follows:
\begin{itemize}
    \setlength\itemsep{.2mm}
    \item How the spectra is produced: DM decay (1) or DM annihilation (2)
    \item The particle type: $\nu_e$, $\nu_\mu$, $\nu_\tau$, or $\gamma$.
    \item Which process: ${\rm DM} {\rm DM} \to \nu/\gamma X$ (1) or ${\rm DM} {\rm DM} \to X X$ (2).
    \item The DM mass $M_{\rm DM}$ in GeV.
    \item The $X$ mass $M_{X}$ in GeV. The helicity/polarization of $X$ can be given, with $m= -1,-1/2,0,1/2,1$. If no value is provided $X$ is assumed to be unpolarized.
    \item The decay modes of $X$ with the format being `$\rm BR$,daughter1,daughter2'. The total branching ratio must sum to 1. An optional parameter $C_{m^\prime} - C_{-m^\prime}$ can be passed in order to fix the shape of the box according to Eq.~\eqref{eq:box shapes}. When no argument is given $C_{m^\prime} - C_{-m^\prime}$ is assumed to be 0.
    \item The number of samples from the spectrum.
    \item The path to the csv file where the points are saved. When no input is given, no save is made. The keyword `plot' or 'logplot' can be entered to plot the sampled data linearly or logarithmically.
\end{itemize}
Acceptable daughter particles of $X$ decay are vl, e, mu, tau, h, z, w, ga, u, d, s, c, b, t, and g. The neutrino type is only important for the production of standard neutrinos; especially, the spectrum of tau neutrinos can differ significantly as compared to the spectra of electron and muon neutrinos when all other parameters are identical. The input can be provided manually or via an input file that is passed as an argument.\\
It is noteworthy that the masses of the DM and $X$ particles are not constrained, while the interpolated spectra used for the daughter particles are tabulated up to 10,000 GeV, so the sampler cannot capture the spectra of the daughter particles with energy greater than 10,000 GeV. \\
One disadvantage of these non-standard spectra is, of course, the enlargement of the parameter space. In the typical indirect search for DM, with a fixed channel, only the mass of the DM particle is important. In these non-standard spectra, however, the dimensionality of the parameter space is at least 2, namely the DM mass and the mass of the BSM particle $X$.
%%%%%%%%%%%%%%%%%%%%%
\section{Conclusion}
%%%%%%%%%%%%%%%%%%%%%%
In this study, we have presented a novel approach for neutrino and photon spectra that is largely model-independent. We have developed a user-specified Monte Carlo sampler to efficiently sample these spectra, which can be found in this \href{https://github.com/ajueid/qcd-dm.github.io.git}{github} repository. While previous experimental searches have focused on neutrino and photon spectra with lines (peaks) and continuous spectra, more complex spectra have been largely overlooked. It is important to note that the spectra we have discussed are not constrained by mass, allowing for arbitrarily high or low energy ranges. Furthermore, the inclusion of a two-dimensional parameter space that includes the mass of dark matter and the mass of a BSM particle X, increases the complexity compared to the typical one-dimensional parameter space. This sampler may prove valuable not only for conventional high-energy astroparticle searches, but also for low-energy searches.

%%%%%%%%%%%%%%%%%%%%%%
\section*{Acknowledgements}
%%%%%%%%%%%%%%%%%%%%

R. RdA is supported by PID2020-113644GB-I00 from the Spanish Ministerio de Ciencia e Innovación.

%%%%%%%%%%%%%%%%%%%%%%%%%%
\bibliographystyle{JHEP}
\bibliography{main}
\end{document}